\documentclass[twocolumn,superscriptaddress,amsmath,aps]{revtex4}
\usepackage{amssymb,mathrsfs,bm,color,times}
\usepackage{graphicx,color}
\usepackage{bm}
\usepackage[hypertex]{hyperref}
\usepackage{amsmath}
\usepackage{hyperref}

\newcommand{\be}{\begin{equation}}
\newcommand{\ee}{\end{equation}}
\newcommand{\bea}{\begin{eqnarray}}
\newcommand{\eea}{\end{eqnarray}}
\newcommand{\bsube}{\begin{subequations}}
\newcommand{\esube}{\end{subequations}}

\newcommand{\Eq}[1]{Eq.\,(\ref{#1})}
\newcommand{\Eqs}[1]{Eqs.\,(\ref{#1})}

\newcommand{\dg}{\dagger}

\newcommand{\ra}{\rangle}

\newcommand{\beq}{\begin{equation}}
\newcommand{\eeq}{\end{equation}}
\newcommand{\beqn}{\begin{eqnarray}}
\newcommand{\eeqn}{\end{eqnarray}}

\newcommand{\nl}{\nonumber \\}

\newcommand{\bsub}{\begin{subequations}}
\newcommand{\esub}{\end{subequations}}

\begin{document}
\title{Probing the non-Abelian fusion of a pair of Majorana zero modes}
\author{Jing Bai} 
\affiliation{Center for Joint Quantum Studies and Department of Physics,
School of Science, Tianjin University, Tianjin 300072, China}
\author{Qiongyao Wang}
\affiliation{Center for Joint Quantum Studies and Department of Physics,
School of Science, Tianjin University, Tianjin 300072, China}
\author{Luting Xu}
\affiliation{Center for Joint Quantum Studies and Department of Physics,
School of Science, Tianjin University, Tianjin 300072, China}
\author{Wei Feng}
\email{fwphy@tju.edu.cn}

\affiliation{Center for Joint Quantum Studies and Department of Physics,
School of Science, Tianjin University, Tianjin 300072, China}
\author{Xin-Qi Li}
\email{xinqi.li@tju.edu.cn}
\affiliation{Center for Quantum Physics and Technologies, School of Physical Science and Technology,
Inner Mongolia University, Hohhot 010021, China}
\affiliation{Center for Joint Quantum Studies and Department of Physics,
School of Science, Tianjin University, Tianjin 300072, China}


\begin{abstract}
{\flushleft In this work}
we perform real time simulations for probing
the non-Abelian fusion of a pair of Majorana zero modes (MZMs).
The nontrivial fusion outcomes can be either a vacuum,
or an unpaired fermion, which reflect the underlying non-Abelian statistics.
The two possible outcomes can cause different charge variations
in the nearby probing quantum dot (QD),
while the charge occupation in the dot is detected by a quantum point contact.
In particular, we find that gradual fusion and gradual coupling of the MZMs to the QD
(in nearly adiabatic switching-on limit)
provide a simpler detection scheme than sudden coupling after fusion
to infer the coexistence of two fusion outcomes,
by measuring the occupation probability of the QD.
For the scheme of sudden coupling (after fusion), we propose and analyze
continuous weak measurement for the quantum oscillations of the QD occupancy.
From the power spectrum of the measurement currents,
one can identify the characteristic frequencies
and infer thus the coexistence of the fusion outcomes.
\end{abstract}
\maketitle


{\flushleft\it Introduction}. ---  The nonlocal nature of
the Majorana zero modes (MZMs) and non-Abelian statistics obeyed
provide an elegant paradigm of topological quantum computation
\cite{Kit01,Kit03,Sar08,Ter15,Sar15,Opp20a}.
In the past decade, after great efforts, considerable progress
has been achieved for realizing the MZMs in various experimental platforms.
Yet, the main experimental evidences are largely associated with
the zero-bias conductance peaks
(see the recent review \cite{GHJ22} and references therein),
which cannot ultimately confirm the realization of MZMs
(even a stable quantized conductance cannot also).
Therefore, an essential milestone step is to identify the MZMs
by probing the underlying non-Abelian statistics,
via either braiding or fusion experiments.

Braiding MZMs in real space can result in quantum state evolution in the manifold
of highly degenerate ground states \cite{Fish11,Opp12,Roy19,Han20},
while fusing the MZMs can yield outcomes of
either a vacuum, or an unpaired fermion (resulting in an extra charge)
\cite{Ali16,BNK20,NC22,Leij22,Zut20,Sau23}.
The latter is owing to the fact that
the MZMs essentially realize ``Ising" non-Abelian anyons,
which obey a particularly simple fusion rule as \cite{Ali16,BNK20}
\bea\label{fusion}
\gamma\times \gamma= I + \psi   \,.
\eea
This means that a pair of MZMs can either annihilate
or combine into a fermion $\psi$.
These two ``fusion channels" correspond to the regular fermion being empty or filled.
The presence of multiple fusion channels is essentially related to
non-Abelian statistics (actually is commonly used to define non-Abelian anyons).
More specifically, there exist two types of fusion design \cite{Ali16,BNK20}.
The ``trivial" fusion corresponds to the fused pair of MZMs
with a defined parity within the same pair.
In this case, the outcome is deterministic;
it leads to unchanged parity with no extra charge.
Of more interest is the case of ``nontrivial" fusion,
where the fused pair of MZMs are from different pairs
with parities (e.g. even parity) being defined in advance.
In this case, the fusion yields probabilistic outcomes as shown above.

While directly probing non-Abelian statistics of MZMs is a milestone
towards topological quantum computation,
probing fusion should be simpler than demonstrating braiding \cite{Ali16,BNK20}.
However, so far there is not yet report of nontrivial fusion experiment \cite{NC22}.
The basic idea of probing fusion is bringing a pair of MZMs together
to remove energy degeneracy
between the two possible outcomes of fusion (i.e., $I$ and $\psi$),
owing to overlap of the two MZMs.
Then, a measurement to distinguish the fermion parity
can reveal the stochastic result being $I$ or $\psi$, with equal probability.
This type of nontrivial fusion demonstration is actually
equivalent to demonstrating the underlying non-Abelian statistics \cite{Ali16,BNK20}.

\begin{figure}
  \centering
  \includegraphics[scale=1.3]{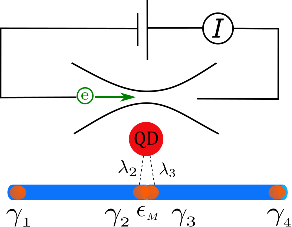}\\
  \caption{
Schematic setup of probing the non-Abelian fusion of a pair of MZMs,
say, $\gamma_2$ and $\gamma_3$, from different Majorana pairs
with parities (e.g. even parity) being defined in advance.
Based on the fusion rule $\gamma\times \gamma= I + \psi$,
the fused MZMs would yield probabilistic outcomes of vacuum $I$ and a regular fermion $\psi$.
A quantum dot is introduced to couple to the fusing MZMs
for probing the fusion outcomes, and a nearby point-contact detector
is introduced to detect the charge occupation of the quantum dot.      }
\end{figure}

In practice, demonstrating nontrivial fusion of MZMs should require
preparation of initial pair states of MZMs with definite fermion parities
and nonadiabatic moving when bringing the MZMs together to fuse.
In this work, along the line proposed in Ref.\ \cite{NC22}
(as schematically shown here in Fig.\ 1),
we consider fusing a pair of MZMs from
two topological superconducting (TSC) wires
(each wire accommodating two MZMs at the ends).
This model setup can correspond to the platform of
mini-gate controlled planar Josephson junctions \cite{RenN19,RenP19}.
The two TSC segments can be formed from a single junction wire,
by making them separated by a topologically trivial segment, via gating technology.
A quantum dot is introduced to couple to the central part of the coupled wires,
for use of probing the fusion outcomes
when bringing the MZMs together to fuse at the central part.
Moreover, a nearby point-contact (PC) detector
is introduced to detect the charge occupation of the quantum dot.
All the ingredients in this proposal
are within the reach of nowadays state-of-the-art experiments \cite{NC22}.
For fusion experiments based on this proposal,
one may encounter some practical complexities,
such as the interplay of charge fluctuations in the quantum dot
caused by the two fusion outcomes, which is relevant to
the control of the dot energy level and its coupling strengths to the MZMs,
and the effect on the dot occupation pattern
caused by the speed of fusion and coupling of the MZMs to the quantum dot.
Detection schemes accounting for these issues will be analyzed in this work,
and are expected to be useful for future experiments.

\vspace{0.3cm}
{\flushleft\it Setup and Basic Consideration}.---
The setup proposed in Ref.\ \cite{NC22} can be modeled as Fig.\ 1,
where the two TSC quantum wires are formed by interrupting
a single TSC wire at the center via mini-gate-voltage control.
For each TSC wire, a pair of MZMs are emergent at the ends,
i.e., ($\gamma_1, \gamma_2$) in the left wire
and ($\gamma_3, \gamma_4$) in the right wire.
The coupling between the central modes $\gamma_2$ and $\gamma_3$
is described by $H'_{M}=i\epsilon_M \gamma_2 \gamma_3$,
with the coupling energy $\epsilon_M$ changeable
when $\gamma_2$ and $\gamma_3$ are separated away in space by mini-gate-voltage control.

Most naturally, one can combine ($\gamma_1,\gamma_2$)
as a regular fermion $f_{12}$ with occupation $n_{12}=0$ or 1,
and ($\gamma_3,\gamma_4$)
as another regular fermion $f_{34}$ with occupation $n_{34}=0$ or 1.
For fusion experiment, one can prepare the specific
initial state $|0_{12}0_{34}\ra$ as proposed in Ref.\ \cite{NC22}.
That is, by means of mini-gate-voltage control,
move $\gamma_2$ and $\gamma_3$ to the ends of the two wires,
close to $\gamma_1$ and $\gamma_4$, respectively;
then, empty the possible occupations of the regular fermions $f_{12}$ and $f_{34}$
by introducing tunnel-coupled side quantum dots and modulating the dot energies
(while the quantum dots are also tunnel-coupled to outside reservoirs).
Starting with $|0_{12}0_{34}\ra$,
consider simultaneously moving $\gamma_2$ and $\gamma_3$
from the two terminal sides
back to the central part to fuse
(to couple each other such that $\epsilon_M\neq 0$), as shown in Fig.\ 1.
For the final state, in the representation of $n_{12}$ and $n_{34}$ occupations,
it is still $|0_{12}0_{34}\ra$.
However, in the representation of $n_{23}$ and $n_{14}$,
i.e., the occupations of the regular fermions $f_{23}$ and $f_{14}$
associated with the Majorana pairs ($\gamma_2,\gamma_3$) and ($\gamma_1,\gamma_4$),
we can reexpress this state as (for derivation of this transformation rule,
or the so-called {\it fusion rule}, see Appendix A)
\bea\label{FR}
|0_{12}0_{34}\ra
= \frac{1}{\sqrt{2}} (|0_{23}0_{14}\ra +i |1_{23}1_{14}\ra)   \,.
\eea
We find that, in the new representation,
the occupation of the $f_{23}$ fermion can be empty or occupied,
i.e., $|0_{23}\ra$ or $|1_{23}\ra$.
This is nothing but the two possible outcomes $I$ and $\psi$ of
the nontrivial fusion of Ising anyons, as shown by \Eq{fusion}.

The fusion coupling between the Majorana modes $\gamma_2$ and $\gamma_3$
would lift the energy degeneracy of the states $|0_{23}\ra$ and $|1_{23}\ra$,
thus allowing to identify the fusion outcomes $I$ and $\psi$.
Following Ref.\ \cite{NC22},
we consider to introduce a nearby quantum dot (QD)
to couple to the central segment of the quantum wire, as shown in Fig.\ 1,
where the Majorana modes $\gamma_2$ and $\gamma_3$ are located.
The QD is assumed to have a single relevant energy level,
described by $H_D=\epsilon_D d^{\dagger}d$.
We thus expect different charge occupation patterns of the QD,
for the different fusion outcomes $I$ and $\psi$.
In this context, it would be more convenient to describe the coupling
between $(\gamma_2,\gamma_3)$ and the QD
using the regular fermion $f_{23}$ picture, as follows
\bea
H'_{DF} = (\lambda_N d^{\dagger} f_{23}
+ \lambda_A d^{\dagger} f_{23}^{\dagger})  + {\rm h.c.}  \,.
\eea
Here we have used the definition $f_{23}=(\gamma_2+i\gamma_3)/2$.
Physically, the first term
describes the usual normal tunneling process
and the second term describes the Andreev process
owing to Cooper pair splitting and recombination.
The coupling amplitudes
are associated with the couplings of
$\gamma_2$ and $\gamma_3$ to the QD,
say, $\lambda_2$ and $\lambda_3$ as shown in Fig.\ 1,
as $\lambda_{N,A}=\lambda_2 \pm i\lambda_3$.

Also following the proposal of  Ref.\ \cite{NC22},
the charge fluctuation in the quantum dot (occupied or unoccupied)
is measured by a point-contact (PC) detector, as schematically shown in Fig.\ 1.
PC detectors with sensitivity at single electron level have been experimentally
demonstrated and broadly applied in practice \cite{Ens06,Fuj06,Guo13,Guo15}.
Actually, the measurement dynamics of a charge qubit by a PC detector
has been a long standing theoretical problem
and has attracted intensive interest
in the community of quantum and mesoscopic physics \cite{Kor01,Mil01,Li05,Opp20}.
In this work, we will perform real-time simulations for
probing the non-Abelian fusion of a pair of MZMs.
In particular, within the scheme of continuous quantum weak measurement,
we will carry out results of individual quantum trajectories
and power spectrum of the measurement currents.
The characteristic frequencies in the power spectrum
indicate the quantum oscillations of charge transfer
associated with the two outcomes of fusion.
The coexistence of two characteristic frequencies should be
a promising evidence for the non-Abelian fusion of MZMs.

\vspace{0.3cm}
{\flushleft\it QD occupations caused by the two fusion outcomes}.---
Let us consider first the detection scheme of switching on the coupling
of the QD to the Majorana modes $\gamma_2$ and $\gamma_3$,
after their fusion from the deterministic initial state $|0_{12}0_{34}\ra$.
This can be realized by initially
setting the QD energy level much higher than
the final coupling energy $\epsilon_M$ between $\gamma_2$ and $\gamma_3$.
Then, during the moving and fusion process,
the QD level is effectively decoupled with $\gamma_2$ and $\gamma_3$,
owing to the large mismatch of energies.
After fusion, switch on the coupling by lowering the QD energy level
such that $\epsilon_D=\epsilon_M$.

For this scheme, the time dependent charge occupation in the dot
is shown in Fig.\ 2 (the result of the green curve).
To understand this result, we should notice the coexistence of
two channels of charge transfer oscillations
between the QD and the MZMs $\gamma_2$ and $\gamma_3$.
One channel is governed by normal tunneling
between the states $|1_{23}0_d\ra$ and $|0_{23}1_d\ra$,
resulting in a quantum oscillation state as
$\alpha_N(t) |1_{23}0_d\ra + \beta_N(t)|0_{23}1_d\ra$,
with the dot occupation probability $p_d^{(N)}(t)=|\beta_N(t)|^2$
plotted in Fig.\ 2 by the full Rabi-type oscillating red curve.
The other channel is governed by the Andreev process
between $|0_{23}0_d\ra$ and $|1_{23}1_d\ra$,
resulting in a quantum oscillation given by
$\alpha_A(t) |0_{23}0_d\ra + \beta_A(t)|1_{23}1_d\ra$,
with the dot occupation probability $p_d^{(A)}(t)=|\beta_A(t)|^2$
plotted in Fig.\ 2 by the smaller amplitude blue curve.
These two channels are independent to each other.
Thus the electron occupation in the dot is simply
an equal probability weighted sum, based on \Eq{FR}, as
\bea
p_d(t) = \left[ |\beta_N(t)|^2 + |\beta_A(t)|^2 \right]/2  \,.
\eea
Actually, the quantum oscillations in the two channels
have simple analytic solutions,
with dot occupation probabilities given by
\bea\label{Rabi}
|\beta_{N,A}(t)|^2
=   \left( |\lambda_{N,A}|^2 / \Omega_{N,A}^2  \right)
\sin^2 \left( \Omega_{N,A} t \right)  \,,
\eea
where $\Omega_{N,A} = \sqrt{ \Delta_{N,A}^2+ |\lambda_{N,A}|^2 }$
and $\Delta_{N,A}=|\epsilon_D\mp\epsilon_M|/2$.
We then understand that, when $\epsilon_D \simeq \epsilon_M >> |\lambda_{N,A}|$,
the quantum oscillation associated with the fusion outcome $\psi$ is dominant,
while the Andreev process following the $I$ outcome is largely suppressed.
However, viewing that the coupling energy $\epsilon_M$
between $\gamma_2$ and $\gamma_3$ is small
(might be comparable with $\lambda_2$ and $\lambda_3$ in practice, see Fig.\ 1),
one may encounter the complexity that both channels coexist
during detection of the charge variations in the quantum dot,
having thus the result as shown in Fig.\ 2 by the green curve.

\begin{figure}
  \centering
  \includegraphics[scale=0.52]{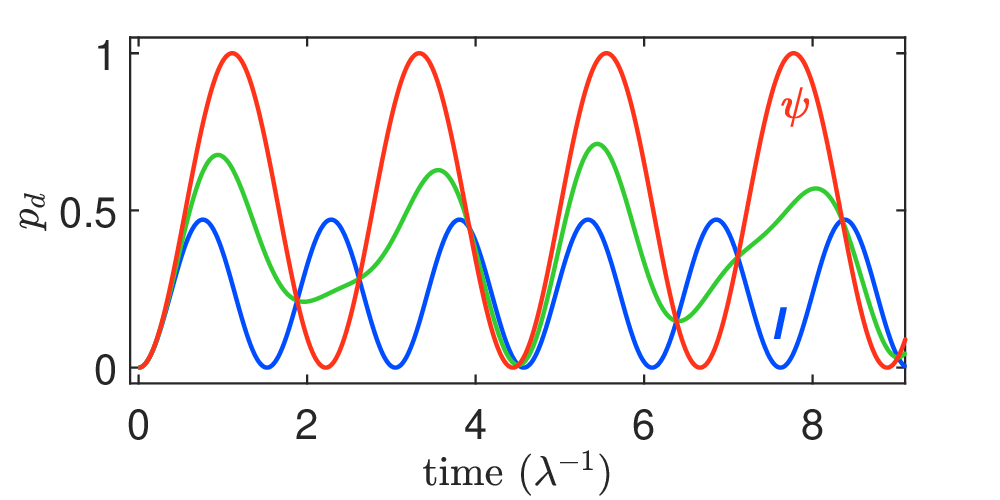}\\
  \caption{
Dot occupation probability $p_d$ (green curve)
associated with the detection scheme of sudden coupling
of the MZMs after fusion to the probing QD.
According to the fusion rule, \Eq{FR},
$p_d$ is half of the sum of $p_d^{(N)}$ (red curve, labeled by $\psi$)
and $p_d^{(A)}$ (blue curve, labeled by $I$).
The meaning of $p_d^{(N)}$ and $p_d^{(A)}$ is referred to the main text.
Parameters are assumed as $\lambda_2=\lambda_3=\lambda=1$,
and $\epsilon_D=\epsilon_M=1.5$.
In this work, we use the arbitrary system of units,
taking $\lambda$ as the unit of energy
and $\lambda^{-1}$ as the unit of time.     }
\end{figure}

Next, let us consider an alternative detection scheme of gradually coupling
the QD with the Majorana modes $\gamma_2$ and $\gamma_3$.
The initial state preparation and moving of the Majorana modes $\gamma_2$ and $\gamma_3$
are the same as above (the first scheme).
However, we consider now initially setting the QD level
in resonance with the final Majorana coupling energy,
i.e., $\epsilon_D=\epsilon_M$.
Then, in this scheme, modulation of the QD energy level
after Majorana fusion is not needed.
When $\gamma_2$ and $\gamma_3$ are somehow slowly brought
to close to each other, they also couple to the QD gradually.
We may model the gradual coupling as follows
\bea\label{coup-model}
& \lambda_2=\lambda_3 = \lambda \left[ \frac{t}{\tau}\Theta(1-\frac{t}{\tau})
+\Theta(\frac{t}{\tau}-1) \right]   \,,    \nl
& \epsilon_{23} = \epsilon_M \left[ \frac{t}{\tau}\Theta(1-\frac{t}{\tau})
+\Theta(\frac{t}{\tau}-1) \right]   \,,
\eea
where $\Theta(\cdots)$ is the step function.
In this simple model, $\tau$ is introduced to characterize
the speed of moving $\gamma_2$ and $\gamma_3$.
Here we assume that the coupling of $\gamma_2$ and $\gamma_3$
to the QD (nonzero $\lambda_2$ and $\lambda_3$) and the coupling
between them (nonzero $\epsilon_{23}$) are started at same time.
We also use it as the initial time for latter state evolution,
which is associated with the fusion and detection.
Before $\gamma_2$, $\gamma_3$ and the QD start to couple to each other,
the moving of $\gamma_2$ and $\gamma_3$
can be performed at different speed (faster speed).
However, the moving should satisfy the adiabatic condition
determined by the energy gap of the TSC wire,
which is much larger than the coupling energies
$\epsilon_M$, $\lambda_2$ and $\lambda_3$, and the dot energy $\epsilon_D$.
We may remark that the modeling, in terms of \Eq{coup-model},
might not be very accurate,
but it captures the underlying physics
and can predict valid behavior of
electron occupation in the quantum dot,
in comparison with
more accurate simulation based on the more realistic lattice model.

\begin{figure}
  \centering
  \includegraphics[scale=0.52]{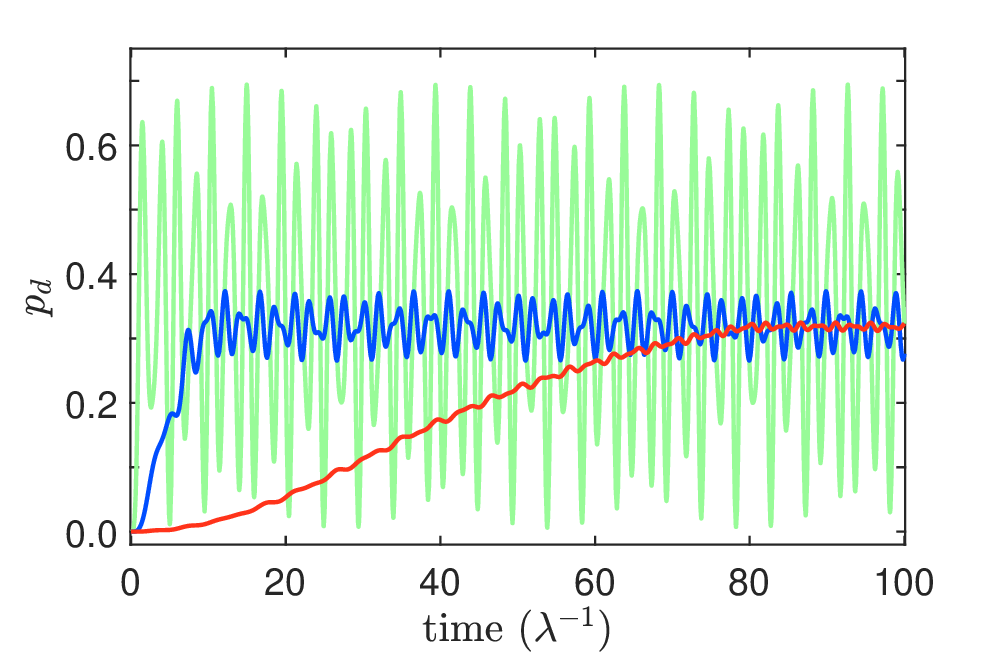}\\
  \caption{
Dot occupation probability $p_d$ associated with the detection scheme of
gradual fusion and coupling to the probing QD, as modeled by \Eq{coup-model}.
Results of different coupling speeds are shown for
$\tau=1$ (green curve), 10 (blue curve), and 80 (red curve), respectively.
Prominent feature is that the quantum oscillations
tend to disappear when decreasing the speed of coupling,
i.e., when approaching the adiabatic limit of switching on the coupling.
Parameters are the same as assumed in Fig.\ 2.        }
\end{figure}

In Fig.\ 3 we show results of different coupling speeds,
which are characterized by the parameter $\tau$.
For fast coupling, the result (green curve) is similar to that shown in Fig.\ 2.
However, if decreasing the speed of coupling,
the charge occupation pattern in the quantum dot becomes different.
The most prominent feature is that the quantum oscillations
tend to disappear (see the blue and red curves in Fig.\ 2).
This can be understood as follows.
In this scheme of fusion and detection,
there exist also two charge transfer channels
associated with, respectively, the fusion outcomes $I$ and $\psi$.
However, in the limit of adiabatically switching on the coupling,
in each channel the state will largely follow an instantaneous eigenstate.
For instance, for the $\psi$-related dominant channel
(governed by the normal tunneling
between the states $|1_{23}0_d\ra$ and $|0_{23}1_d\ra$),
the state can be expressed also as
$\alpha_N(t) |1_{23}0_d\ra + \beta_N(t)|0_{23}1_d\ra$.
Yet, in the adiabatic limit, the superposition coefficients
in the instantaneous eigenstate
do not reveal the feature of Rabi-type quantum oscillations.
Actually, we can easily obtain $|\beta_N(t\ge \tau)|^2 = p^{(N)}_d = 1/2$.
Similarly, for the outcome-$I$-related channel,
the instantaneous eigenstate can be expressed as
$\alpha_A(t) |0_{23}0_d\ra + \beta_A(t)|1_{23}1_d\ra$.
In the adiabatic limit, $|\beta_A(t)|^2$ does not oscillate with time
and, when $t\ge \tau$, we obtain
\bea
p^{(A)}_d = |\beta_A|^2
=\frac{(\Omega_A-\epsilon_M)^2}{(\Omega_A-\epsilon_M)^2 + 2|\lambda|^2} \,.
\eea
Here, $\Omega_A = \sqrt{\epsilon^2_M + 2|\lambda|^2}$,
under the conditions $\epsilon_D=\epsilon_M$ and $\lambda_2=\lambda_3=\lambda$.
Based on the fusion rule \Eq{FR}, we expect the overall occupation probability
of an electron in the QD to be $p_d=(p^{(N)}_d + p^{(A)}_d)/2$.
Indeed, this is the asymptotic result observed in Fig.\ 3, in the adiabatic limit.

The QD occupations predicted in Figs.\ 2 and 3 can be measured
by using a charge-sensitive PC-detector as shown in Fig.\ 1.
The standard method is performing the so-called single shot
projective measurement to infer the QD being occupied or not by an electron.
After a large number of ensemble measurements, the occupation probability can be obtained.
However, the result in  Fig.\ 2
(the overall pattern plotted by the green curve)
does not very directly reveal the coexistence of the two fusion outcomes.
Also, measuring this oscillation pattern and the subsequent analyzing
will be more complicated than handling the result
from the adiabatic coupling as shown in Fig.\ 3.
That is, importantly,
measuring the final single constant occupation probability
is much simpler than measuring the oscillation pattern in Fig.\ 2,
while the result can more directly inform us
the coexistence of the two fusion outcomes,
by using the formula $p_d=(p^{(N)}_d + p^{(A)}_d)/2$
and the result $p^{(N)}_d =0.5$.
Therefore, the second detection scheme proposed above
is expected to be useful in practice,
by adiabatically switching on the probe coupling.  \\
\\
{\it Continuous weak measurements}.---
Besides the usual strong (projective) measurements,
as discussed above for measuring the QD occupation probability,
continuous quantum weak measurement
is an interesting and different type of choice,
suitable in particular for measuring quantum oscillations.
For instance, the problem of continuous weak measurement
of charge qubit oscillations by a PC detector
has attracted strong interest for intensive studies \cite{Kor01,Mil01,Li05}.
In the following, we consider continuous weak measurement
for the quantum oscillations displayed in Fig.\ 2 (by the green curve).
Specifically, for the setup shown in Fig.\ 1,
the PC detector is switched on (applied bias voltage)
from the beginning of state evolution after fusion,
owing to tunnel-coupling with the QD.
For continuous weak measurement, the measurement rate $\kappa$,
see the following \Eqs{Ict} and (\ref{QTE}),
is much smaller than the characteristic energy of the system under measurement,
or, equivalently speaking, the measurement time ($\sim 1/\kappa$)
is much longer than the quantum oscillation period (e.g., in Fig.\ 2). 
Accordingly, the noisy output current in the PC detector
can be expressed as \cite{Kor01,Mil01,Opp20}
\bea\label{Ict}
I_c(t) = n_{d,c}(t) +   \frac{1}{\sqrt{4\kappa}} \, \xi(t)  \,.
\eea
This (rescaled) expression of current
is valid up to a constant factor (with current dimension).
Then, the first term
is simply the quantum average occupation of an electron in the quantum dot,
i.e., $n_{d,c}(t)={\rm Tr}[\hat{n}_d \rho_c(t)]$,
with $\rho_c(t)$ the PC-current-conditioned state of the measured system.
The second term describes the deviation of the real noisy current
from the quantum average occupation $n_{d,c}(t)$,
owing to classical events of
random tunneling of electrons in the PC detector.
The rate parameter $\kappa$ characterizes the measurement strength
and $\xi(t)$ is the Gaussian white noise.
For completeness, we also present here the quantum trajectory (QT) equation
for the conditional state as \cite{Mil01}
\bea\label{QTE}
\dot{\rho}_c = \mathcal{L}\rho_c
    +\sqrt{\kappa}\mathcal{H}[\hat n_d]\rho_c\xi(t)  \,.
\eea
The first deterministic part is given by
$\mathcal{L}\rho_c = -i\left[H , \rho_c \right] + \kappa \mathcal{D}[\hat n_d]\rho_c$,
with the Lindblad superoperator defined as
$\mathcal{D}[x] \rho_c =x \rho_c x^{\dagger}
-\frac{1}{2}\{x^{\dagger}x,\rho_c \}$.
The second noisy term stems from measurement backaction
owing to information gain
in the single realization of continuous weak measurement,
while the superoperator is defined as
$\mathcal{H}[x]\rho_c = x \rho_c+\rho_c x^{\dg}
-\mathrm{Tr}\left[(x+x^{\dg})\rho_c\right]\rho_c$.

\begin{figure}
  \centering
  \includegraphics[scale=0.52]{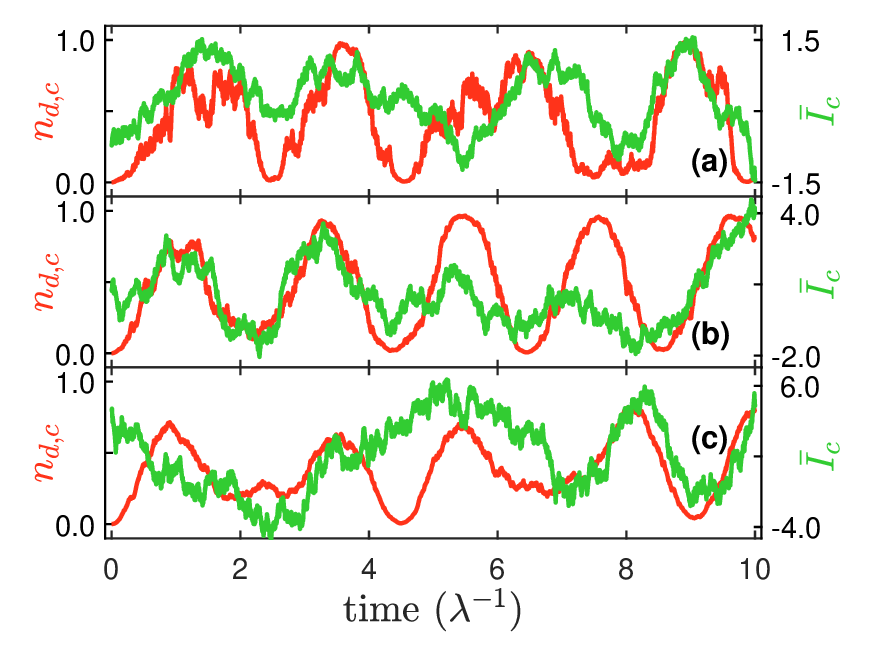}\\
  \caption{
Quantum trajectories of continuous weak measurement,
associated with the detection scheme of
sudden coupling of the MZMs after fusion to the probing QD.
The measurement-current-conditioned occupation $n_{d,c}(t)$
and the current $\bar{I}_c(t)$
(low-pass-filtered with a time window $T=1$)
are shown for measurement strengths
$\kappa=0.8$ in (a), 0.2 in (b), and 0.05 in (c), respectively.
Other parameters are the same as in Fig.\ 2.        }
\end{figure}

Jointly simulating the evolution of \Eqs{Ict} and (\ref{QTE})
we can obtain $\rho_c(t)$, $n_{d,c}(t)$, and $I_c(t)$.
From \Eq{Ict}, we understand that the measurement current
does encode the information of the QD occupation.
However, owning to measurement backaction,
the measurement-current-conditioned occupation $n_{d,c}(t)$ is different from
the occupation probability $p_d(t)$ shown in Fig.\ 2 (in the absence of measurement).
For considerably weak measurement,
$n_{d,c}(t)$ should be quite close to $p_d(t)$,
yet the noisy term in \Eq{Ict} will hide the informational term,
thus preventing us inferring the dot occupation.
In contrast, if increasing the measurement strength,
the noisy term will decrease,
but the measurement backaction will make $n_{d,c}(t)$
more seriously deviate from the original result $p_d(t)$.
In Fig.\ 4, we show the results of $n_{d,c}(t)$ and $I_c(t)$,
for a couple of measurement strengths.
In addition to properly choosing the measurement strengths,
we also applied the so-called low-pass-filtering technique.
That is, we averaged the current over a sliding time window of duration $T$,
$\bar{I}_c(t)=\frac{1}{T}\int_{t-T/2}^{t+T/2} dI_c(\tau)$,
which gives a smoothed current for better reflecting the dot occupation.
However, even after making these efforts,
we find that from the noisy output current $I_c(t)$,
it is hard to track the quantum oscillations of the QD occupation shown in Fig.\ 2.

Actually, in continuous weak measurements,
a useful technique is extracting information from
the power spectrum of the measurement currents \cite{Kor01,Mil01,Li05}.
The steady-state current correlation function
is obtained
through the ensemble average
$S_I(\tau) = {\rm E}[I_c(t+\tau)I_c(t)]- {\rm E}[I_c(t+\tau)] {\rm E}[I_c(t)]$,
at long time limit (large $t$ limit for achieving steady state).
From the power spectrum,
$S_I(\omega)=\int^{\infty}_{-\infty} d\tau  S_I(\tau) e^{i\omega\tau}$,
one can identify the characteristic frequencies
and infer thus quantum coherent oscillations inside the system under measurement.
For the problem under study,
the goal is to identify the quantum oscillations shown in Fig.\ 2,
which are associated with the two fusion outcomes.
Based on the result of \Eq{Ict}, it can be proved \cite{Kor01,Mil01} that
\bea\label{SI}
S_I(\tau)
&=& S_d(\tau)+\frac{1}{4 \kappa} \delta(\tau)-\frac{1}{4} \,.
\eea
In this result, the correlation function of the dot occupation is given by
$ S_d(\tau)={\rm E}[n_{d,c}(t+\tau) n_{d,c}(t)]
= {\rm Tr} [  {\hat{n}_d} e^{\mathcal{L}|\tau|} (\hat{n}_d \rho_{st})]$,
with $\rho_{st}$ the reduced density matrix of steady state.
Then, we know the structure of the current power spectrum as
$S_I(\omega)=S_0 + S_d(\omega)$,
with $S_0$ the frequency-free background noise
and $S_d(\omega)$ the information-contained part.
Based on the master equation
$\dot{\rho}={\cal L}\rho$,
which is the ensemble average of \Eq{QTE},
using the so-called quantum regression theorem we obtain
\bea\label{Laplace}
&& S_d^{(j)}(\omega)  \nl
&=& \frac{2\kappa |\lambda_j|^2 (16\Delta_j^2+\kappa^2+4\omega^2)}
{\omega^2(16\Omega_j^2+\kappa^2-4\omega^2)^2+16\kappa^2(2|\lambda_j|^2-\omega^2)^2} .
\eea
Here we use $j=N,A$ to denote the two charge transfer channels,
say, the Andreev process and normal tunneling,
with coupling amplitudes $\lambda_{N,A}=\lambda_2 \pm i\lambda_3$.
Since the two channels are independent,
we obtained the above results independently for each process.
The overall spectrum $S_d(\omega)$ is the weight-averaged sum of
$S_d^{(N)}(\omega)$ and $S_d^{(A)}(\omega)$,
i.e., $S_d(\omega)=[S_d^{(N)}(\omega)+S_d^{(A)}(\omega)]/2$,
owing to the fusion rule of \Eq{FR}.
Moreover, under the condition of weak-coupling measurement,
we can further approximate the result as
\bea\label{Lor}
&& S_d^{(j)}(\omega)
\simeq \frac{\Delta_j^2}{4\Omega_j^2}
\frac{\kappa R_j/2}{\omega^2 + (\kappa R_j/2)^2} \nl
&& ~ + \frac{|\lambda_j|^2}{8\Omega_j^2}
\frac{\frac{\kappa}{2} (1-\frac{R_j}{2})}
{(\omega-2\Omega_j)^2+[\frac{\kappa}{2} (1-\frac{R_j}{2})]^2 }  \,.
\eea
Here we have introduced $R_j \equiv |\lambda_j|^2/\Omega_j^2$.
From this standard Lorentzian form, one can extract
the characteristic frequencies $\Omega_N$ and $\Omega_A$,
which reflect in essence the quantum oscillations given by \Eq{Rabi}.

\begin{figure}
  \centering
  \includegraphics[scale=0.57]{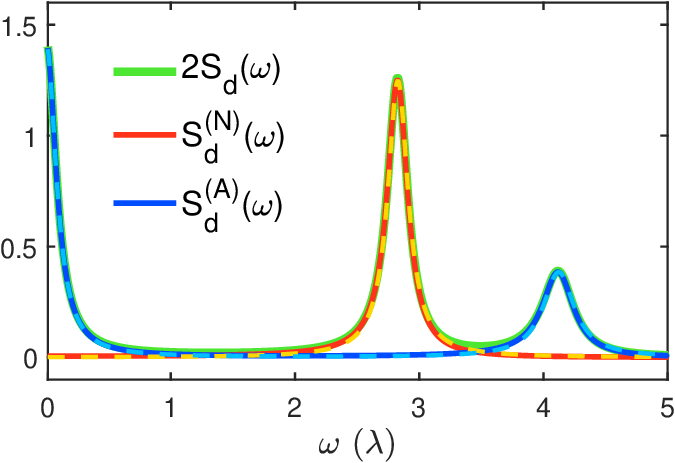}\\
  \caption{
Characteristic frequency spectrum, $S_d(\omega)$,
the Fourier transform of the QD-occupation correlation function,
from the output currents of continuous weak measurement
associated with the detection scheme of sudden coupling of the MZMs after fusion to the probing QD.
Theoretically, owing to the fusion rule \Eq{FR},
$S_d(\omega)=[S_d^{(N)}(\omega)+S_d^{(A)}(\omega)]/2$,
where $S_d^{(N)}(\omega)$ and $S_d^{(A)}(\omega)$
are the spectrums related to the fusion outcomes $\psi$ and $I$.
Exact results of \Eq{Laplace}
are displayed by the solid-red and solid-blue curves,
while approximate results of the Lorentzian form \Eq{Lor}
are plotted by the dashed-yellow and dashed-blue curves.
$S_d(\omega)$ is from numerically solving the full master equation,
which includes the two fusion outcomes caused charge transfer channels.
Satisfactory agreement between all these results is demonstrated.
The coexistence of two Lorentzian peaks
(at $2\Omega_N$ and $2\Omega_A$) in $S_d(\omega)$
indicates the appearance of the two fusion outcomes, predicted by \Eq{fusion}.
Measurement strength $\kappa=0.4$ is assumed,
while other parameters are the same as in Fig.\ 2.     }
\end{figure}

In Fig.\ 5, we plot the result of $2 S_d(\omega)$
from numerically solving the full master equation,
which includes the two charger transfer channels.
We find that it is indeed the sum of $S_d^{N}(\omega)$ and $S_d^{A}(\omega)$,
while in this plot, for the purpose of self-consistence verification,
we use their analytic solutions given by \Eq{Laplace}.
We also compare the results with the approximate Lorentzian form solutions
and find satisfactory agreement.
Very importantly,
the coexistence of two Lorentzian peaks
(at $2\Omega_N$ and $2\Omega_A$) in $S_d(\omega)$
simply indicates the appearance of the two fusion outcomes, predicted by \Eq{fusion}.

We may remark that, within the scheme of continuous quantum weak measurement,
from its output current power spectrum $S_I(\omega)$
to infer the intrinsic quantum oscillations
is a very useful scheme,
which is much simpler than the ensemble single shot
projective measurements of the dot occupation,
in order to obtain the probabilities as shown in Figs. 2 and 3.
This type of technique has been analyzed in detail
in the context of charge-qubit measurements \cite{Kor01,Mil01,Li05}.
The present proposal is an extension along this line,
hopefully to be employed to identify the non-Abelian fusion of MZMs,
through the two different characteristic frequencies of quantum oscillations,
which are associated with the two fusion outcomes. \\
\\
{\flushleft\it Summary and Discussion}.---
We have analyzed two schemes of detecting the nontrivial fusion of a pair of MZMs.
The two possible stochastic fusion outcomes reflect
the non-Abalian statistics nature of the MZMs,
whose experimental demonstration will be a milestone
for ultimately identifying the MZMs
and paving the way to topological quantum computation.
One scheme, the most natural choice,
is to switch on sudden coupling of the fused MZMs to the probing QD,
with the subsequent QD oscillating occupation
being monitored by a PC detection in terms of continuous weak measurement.
From the power spectrum of the measurement currents,
one can identify two characteristic frequencies of quantum oscillations
and infer thus the two fusion outcomes of the pair of MZMs.
The other scheme is to switch on, almost adiabatically,
gradual fusion coupling between the MZMs
and their coupling to the probing QD.
This type of detection scheme will result in the QD occupation not oscillating with time,
thus allowing a simpler way to measure the single value of the QD occupation probability
and using it to infer the two outcomes of nontrivial fusion.
We expect that both detection schemes analyzed in this work
can be useful for future fusion experiments.

Before finishing the work, we may make some remarks on issues
of the quantum measurement and some practical complexities.
For the quantum measurement of the QD occupation probabilities
shown in Figs.\ 2 and 3, it should be the well established
single-shot projective measurement \cite{Ens06,Fuj06,Guo13,Guo15}.
For this type of measurement,
the process of wavefunction collapse is usually not described,
despite that physically it is still gradual,
but the collapse time (measurement time)
is much shorter than the characteristic evolution time of the system.
The second type measurement analyzed in this work is,
after fusion of $\gamma_2$ and $\gamma_3$ (see Fig.\ 1)
and modulating the QD level such that $\epsilon_D=\epsilon_M$,
switching on the continuous weak measurement by the PC detector
(applying a bias voltage cross it), to get the steady state current power spectrum.
Notice that this type of measurement is also well established,
see Refs.\ \cite{Kor01,Mil01,Li05,Opp20} and the references therein
(especially, the recent work \cite{Opp20}
for the measurement of Majorana qubts,
via the spectrum of the PC detector current).
We may point out that, even for the nontrivial fusion,
each time one can obtain only $S_d^{(N)}(\omega)$ or $S_d^{(A)}(\omega)$.
Ensemble average of large number of results
will give the spectrum $S_d(\omega)$ shown in Fig.\ 5.

We also point out that,
the fusion of $\gamma_2$ and $\gamma_3$ and the fusion outcome probing
are within the subspace of low energy states.
The energy scales are determined by the Majorana coupling energy $\epsilon_M$,
the QD energy level $\epsilon_D$, and the coupling strengths $\lambda_2$ and $\lambda_3$.
Either the {\it slow} or relatively {\it fast} probe coupling
associated with the results in Fig.\ 3
is with respect to these energies,
which are much smaller than the gap of the TSC wire.
Thus the nonadiabatic transition
to the excited Bogoliubov quasi-particle states can be ignored.
For the probe coupling (with the QD)
after the fusion of $\gamma_2$ and $\gamma_3$,
with the result shown in Fig.\ 2,
the QD level $\epsilon_D$ is modulated from large off-resonance
to resonance with the Majorana coupling energy $\epsilon_M$,
i.e., $\epsilon_D=\epsilon_M$.
In this case, the nonadiabatic transition of MZMs
to higher-energy Bogoliubov states becomes even less relevant,
viewing that the coupling strengths $\lambda_2$ and $\lambda_3$ are also weak.
However, for Majorana moving along the TSC wire,
nonadiabatic transition may happen,
if the moving speed is not well controlled.
In this case, the impact of nonadiabatic transition
on the nontrivial fusion analyzed in this work
is an interesting problem, which is to be studied in our next work.

We may emphasize that for trivial fusion, which corresponds to preparing
the Majorana pair $\gamma_2$ and $\gamma_3$
in a state with definite fermion parity,
the fusion outcome is deterministic, being either $I$ or $\psi$.
Then, for the first probing scheme,
say, adiabatically switching on the probe coupling,
the QD occupation probability is either $p_d=p^{(N)}_d$ or $p_d=p^{(A)}_d$,
which is different from the result of nontrivial fusion,
$p_d=(p^{(N)}_d + p^{(A)}_d)/2$.
From the quite different occupation probability,
one can infer the fusion being trivial or nontrivial.
For the second probing scheme,
say, using the output current power spectrum
of continuous weak measurement, as shown in Fig.\ 5,
there will be two spectral peaks
for nontrivial fusion of $\gamma_2$ and $\gamma_3$,
while for trivial fusion, only one spectral peak appears.
Also, the peak heights are different.

Finally, we mention that a possible complexity in real system may be caused by
the so-called partially separated Andreev bound states (ps-ABSs) \cite{Vui19}.
For the setup schematically shown in Fig.\ 1, in the case of ps-ABSs,
the Majorana $\gamma_1$ would couple to $\gamma_2$ and $\gamma_3$,
and the Majorana $\gamma_4$ would couple to $\gamma_3$ and $\gamma_2$.
Then, in terms of regular fermion picture,
$f_{23}$ and $f_{14}$ will couple to each other.
We know that the nontrivial fusion of $\gamma_2$ and $\gamma_3$
has two outcomes $I$ and $\psi$,
which correspond to the occupation of the $f_{23}$ fermion $n_{23}=0$ and $1$,
and result in two different charge transfer channels with the probing QD.
Therefore, the coupling between $f_{23}$ and $f_{14}$
would ``swap" the occupations of $n_{23}=0$ and $1$,
which would complicate the probing dynamics
associated with the two outcomes $I$ and $\psi$,
and hinder us to identify the fusion outcomes.
In practice, this complexity should be avoided.

\vspace{0.3cm}
{\flushleft\it Acknowledgements.}---
This work was supported by the NNSF of China
(Grants No.\ 11974011 and No.\ 11904261).

\appendix
\section{Derivation of the Fusion Rule}

In this Appendix let us prove
the following transformation rule (fusion rule):
\bea 
|0_{12}0_{34}\ra
= \frac{1}{\sqrt{2}} (|0_{23}0_{14}\ra + i |1_{23}1_{14}\ra)  \,.  \nonumber
\eea
Under the constraint of fermion parity, in general,
we may first construct the transformation ansatz as
$|0_{12}0_{34}\ra = a |0_{23}0_{14}\ra + b |1_{23}1_{14}\ra$.
Then, we express the operator $f_{12}=(\gamma_1+i\gamma_2)/2$
in terms of the regular fermion operators $f_{14}$ and $f_{23}$ as
\bea
f_{12} = \frac{1}{2} \left( f_{14}+f^{\dagger}_{14}
+i f_{23}+i f^{\dagger}_{23}  \right)  \,.
\eea
This result is obtained by simply associating the Majorana fermions
$\gamma_1$ and $\gamma_4$ with the regular fermion $f_{14}$,
and $\gamma_2$ and $\gamma_3$ with $f_{23}$, respectively.
Thus we have $\gamma_1=f_{14}+f^{\dagger}_{14}$
and $\gamma_2=f_{23}+f^{\dagger}_{23}$.
Further, acting the annihilation operator $f_{12}$
on both sides of the ansatz equation, we have
\bea
&& 0 = a (|0_{23}1_{14}\ra +i |1_{23}0_{14}\ra ) \nl
&& ~~~ + b ( i|0_{23}1_{14}\ra - |1_{23}0_{14}\ra)   \,.
\eea
During the algebra, one should notice the difference of a minus sign
between $f_{23}|1_{23}1_{14}\ra=|0_{23}1_{14}\ra$
and $f_{14}|1_{23}1_{14}\ra=-|1_{23}0_{14}\ra$.
From this result, we obtain $ia-b=0$ and $a=1/\sqrt{2}$,
which finishes the proof of the above formula of transformation.
Applying the same method outlined above,
one can carry out all the transformation formulas between
the two sets of basis states, $|n_{12} n_{34}\ra$ and $|n_{23} n_{14}\ra$.

\vspace{1.2cm}


\begin{references}

\bibitem{Kit01}
A. Y. Kitaev, {\it Unpaired Majorana fermions in quantum wires}, Phys. Usp.  {\bf 44}, 131 (2001).
\bibitem{Kit03}
A. Y. Kitaev, {\it Fault-tolerant quantum computation by anyons},
Ann. Phys.  {\bf 303}, 2 (2003).
\bibitem{Sar08}
C. Nayak, S. H. Simon, A. Stern, M. Freedman, and S. D. Sarma,
{\it Non-Abelian anyons and topological quantum computation}, Rev. Mod. Phys.  {\bf 80}, 1083 (2008).
\bibitem{Ter15}
B. M. Terhal, {\it Quantum error correction for quantum memories},
Rev. Mod. Phys.  {\bf 87}, 307-346 (2015).
\bibitem{Sar15}   
S. D. Sarma, M. Freedman, and C. Nayak,
{\it Majorana zero modes and topological quantum computation},
Quantum Inf.  {\bf 1}, 15001 (2015).
\bibitem{Opp20a}
Y. Oreg and F. von Oppen,
{\it Majorana Zero Modes in Networks of Cooper-pair Boxes:
Topologically Ordered States and Topological Quantum Computation},
Annu. Rev. Condens. Matter Phys.  {\bf 11}, 397 (2020).


\bibitem{GHJ22}
G. Li, S. Zhu, P. Fan, L. Cao, and H.-J. Gao,
{\it Exploring Majorana zero modes in iron-based superconductors},
Chin. Phys. B {\bf 31}, 080301 (2022).



\bibitem{Fish11}
J. Alicea, Y. Oreg, G. Refael, F. von Oppen, and M. Fisher,
{\it Non-Abelian statistics and topological quantum information processing in 1D wire networks},
Nat. Phys.  {\bf 7}, 412 (2011).
\bibitem{Opp12}
B. I. Halperin, Y. Oreg, A. Stern, G. Refael, J. Alicea, and F. von Oppen,
{\it Adiabatic manipulations of Majorana fermions in a three-dimensional
network of quantum wires}, Phys. Rev. B  {\bf 85}, 144501 (2012).
\bibitem{Roy19}
F. Harper, A. Pushp, and R. Roy,
{\it Majorana braiding in realistic nanowire Y-junctions and tuning forks},
Phys. Rev. Research  {\bf 1}, 033207 (2019).
\bibitem{Han20}
C. Tutschku, R. W. Reinthaler, C. Lei, A. H. MacDonald, and E. M. Hankiewicz,
{\it Majorana-based quantum computing in nanowire devices},
Phys. Rev. B  {\bf 102}, 125407 (2020).



\bibitem{Ali16}   
D. Aasen, M. Hell, R. Mishmash, A. Higginbotham, J. Danon, M. Leijnse,
T. Jespersen, J. Folk, C. Marcus, K. Flensberg, and J. Alicea,
{\it Milestones Toward Majorana-Based Quantum Computing},
Phys. Rev. X {\bf 6}, 031016 (2016).
\bibitem{BNK20}
C. W. J. Beenakker,
{\it Search for non-Abelian Majorana braiding statistics in superconductors},
SciPost Phys. Lect. Notes 15 (2020).

\bibitem{NC22}
T. Zhou, M. C. Dartiailh, K. Sardashti, J. E. Han, A. Matos-Abiague,
J. Shabani, and I. \u{Z}uti\'{c},
{\it Fusion of Majorana bound states with mini-gate control in two-dimensional systems},
Nat. Commun. {\bf 13}, 1738 (2022).

\bibitem{Leij22}
R. S. Souto, and M. Leijnse,
{\it Fusion rules in a Majorana single-charge transistor},
SciPost Phys. {\bf 12}, 161 (2022).

\bibitem{Zut20}
T. Zhou, M. C. Dartiailh, W. Mayer, J. E. Han, A. Matos-Abiague, J. Shabani, and I. \u{Z}uti\'{c},
{\it Phase Control of Majorana Bound States in a Topological X Junction},
Phys. Rev. Lett. {\bf 124}, 137001 (2020).

\bibitem{Sau23}
C.-X. Liu, H. Pan, F. Setiawan, M. Wimmer, and J. D. Sau,
{\it Fusion protocol for Majorana modes in coupled quantum dots},
Phys. Rev. B {\bf 108}, 085437 (2023).


\bibitem{RenN19}
 H. Ren, F. Pientka, S. Hart, A. T. Pierce, M. Kosowsky, L. Lunczer, R. Schlereth, B. Scharf, E. M. Hankiewicz, L. W. Molenkamp, B. I. Halperin, and A. Yacoby. {\it Topological Superconductivity in a Phase-Controlled Josephson Junction}, Nature {\bf 569}, 93 (2019).
\bibitem{RenP19}
B. Scharf, F. Pientka, H. Ren, A. Yacoby, and E. M. Hankiewicz, {\it Tuning topological superconductivity in phase-controlled Josephson junctions with Rashba and Dresselhaus spin-orbit coupling},
Phys. Rev. B {\bf 99}, 214503 (2019).





\bibitem{Ens06}
S. Gustavsson, R. Leturcq, B. Simovic, R. Schleser, T. Ihn,
P. Studerus, K. Ensslin, D. C. Driscoll, and A. C. Gossard,
{\it Counting Statistics of Single Electron Transport in a Quantum Dot},
Phys. Rev. Lett. {\bf 96}, 076605 (2006).
\bibitem{Fuj06}
T. Fujisawa, T. Hayashi, R. Tomita, and Y. Hirayama,
{\it Bidirectional Counting of Single Electrons},
Science {\bf 312}, 1634 (2006).
\bibitem{Guo13}
G. Cao, H. O. Li, T. Tu, L. Wang, C. Zhou, M. Xiao, G. C. Guo,
H. W. Jiang, and G. P. Guo,
{\it Ultrafast universal quantum control of a quantum dot charge qubit using Landau-Zener-Stuckelberg interference},
Nat. Commun. {\bf 4}, 1401 (2013).
\bibitem{Guo15}
H. O. Li, G. Cao, G. D. Yu, M. Xiao, G. C. Guo, H. W. Jiang,
and G. P. Guo,
{\it Conditional rotation of two strongly coupled semiconductor charge qubits},
Nat. Commun. {\bf 6}, 7681 (2015).

\bibitem{Kor01}
A. N. Korotkov, {\it Output spectrum of a detector measuring quantum oscillations},
Phys. Rev. B {\bf 63}, 085312 (2001).

\bibitem{Mil01}
H. S. Goan and G. J. Milburn,
{\it Dynamics of a mesoscopic qubit under continuous quantum measurement},
Phys. Rev. B {\bf 64}, 235307 (2001).

\bibitem{Li05}
X. Q. Li, P. Cui, and Y. J. Yan,
{\it Spontaneous Relaxation of a Charge Qubit under Electrical Measurement},
Phys. Rev. Lett. {\bf 94}, 066803 (2005).
\bibitem{Opp20}
J. F. Steiner and F. von Oppen,
{\it Readout of Majorana qubits}, Phys. Rev. Research {\bf 2}, 033255 (2020).


\bibitem{Vui19}
A. Vuik, B. Nijholt, A. R. Akhmerov, and M. Wimmer,
{\it Reproducing topological properties with quasi-Majorana states},
SciPost Phys. {\bf 7}, 061 (2019).  


\end{references}
\end{document}